\documentstyle[aps,manuscript,epsfig]{revtex}

\def\Journal#1#2#3#4{{#1}#2 (#4) #3}
\def\NPB{{Nucl.~Phys.} B}
\def\PLB{{Phys.~Lett.}  B}

\def\PRD{{Phys.~Rev.} D}

\def\CPC{{Comp.~Phys.~Comm.} }
\def\EPJC{{Eur.~Phys.~J.} C}

\newcommand{\pmdg}[2]{{#1} $\pm$ {#2}}

\def\bq{\begin{equation}}
\def\eq{\end{equation}}
\def\ba{\begin{eqnarray}}
\def\ea{\end{eqnarray}}
\def\O{{\cal{O}}}
\def\mboxsc#1{\mbox{\scriptsize #1}}
\def\smin{s_{\mbox{\scriptsize min}}}
\def\ycut{y_{\mbox{\scriptsize cut}}}
\def\sminn{s_{\mbox{\scriptsize min}}^{\mbox{\scriptsize nlo}}}
\def\deltan{\delta^{\mbox{\scriptsize nlo}}}

\begin{document}

\tightenlines

\title{\vskip-3cm{\baselineskip14pt
\centerline{\normalsize MPI-PhT/2000-24\hfill}
\centerline{\normalsize hep-ph/0007172\hfill}
\centerline{\normalsize July 2000\hfill}
}
\vskip1.5cm
Combining QCD Matrix Elements \\at Next-to-Leading Order
with Parton Showers \\ in Electroproduction}
\author{Bj\"orn P\"otter}
\address{Max-Planck-Institut f\"ur Physik \\(Werner-Heisenberg-Institut)\\
F\"ohringer Ring 6, 80805 Munich, Germany}
\maketitle

\begin{abstract}
We present a method to combine next-to-leading order (NLO) 
matrix elements in QCD with leading logarithmic parton showers by
applying a suitably modified version of the phase-space-slicing
method. The method consists of subsuming the NLO corrections into a
scale-dependent phase-space-slicing parameter, which is then
automatically adjusted to cut out the leading order, virtual, soft and
collinear contributions in the matrix element calculation. In this
way a positive NLO weight is obtained, which can be redistributed by 
a parton shower algortihm. As an example, we display the method for
single-jet inclusive cross sections at ${\cal O}(\alpha_s)$ in
electroproduction. We numerically compare the modified version of the
phase-space-slicing method with the standard approach and find very
good agreement on the percent level.
\end{abstract}

\newpage

\section{Introduction}

Much progress has been made in the last years in measuring the
hadronic final state in $eP$-scattering at HERA with high precision
(see \cite{kuhlen} for a recent review). The theoretical tools which
are at hand to describe the hadronic final state are basically fixed
order perturbative calculations, which for most processes are
available at next-to-leading order (NLO), or a combination of leading
order (LO) matrix elements with parton showers (PS), mostly at leading
logarithmic accuracy, which are implemented in event generators (see
\cite{mc} for an overview of available Monte Carlo programs for both
approaches). However, it appears that the theoretical calculations are
a considerable source of errors in determining physical parameters. As
an example, the statistical errors in a recent determination of
$\alpha_s$ from dijet production \cite{al2j} are at the one-percent
level. The systematical and theoretical errors, on the other hand, are
considerably larger and both lie around five percent. Therefore, an
improvement of the theoretical tools is needed.

The fixed higher order and the PS approaches have complimentary
strengths. The fixed higher order calculations reduce uncertainties due to
unphysical renormalization and factorization scale dependences. 
Wide-angle emission of partons, where interference effects between a
large number of diagrams may be important, is described well.
The PS on the other hand allows a description of the cross
section in regions where $\alpha_s$ becomes large, especially in the
region of collinear particle emission, by means of a resummation of
large logarithmic terms. This allows to e.g.\ describe
reasonably well the substructure of jets. In addition, the PS can be
terminated at some small scale $Q_0$, which allows to attach some kind
of hadronization model, as, e.g., the Lund model \cite{lund}, to
describe the non-perturbative region. It is desirable to find a
way of combining the advantages of both approaches into a NLO
event generator. In a complete NLO event generator one certainly would
like to include not only the matrix elements at NLO, but likewise
the PS in next-to-leading log accuracy. Two problems in combining
parton showers and matrix elements can be identified. First, one has
to avoid double counting of events which are included  both in the
matrix element calculations and in the PS. Second, negative weights
can occur in the calculation of the matrix elements at NLO. Although
they are not a principle problem, negative weigths can make it in
practice difficult to obtain numerically stable results, when they are
used as a starting point for the PS.

Basically two strategies can be adopted to combine matrix elements and PS's. 
Either the phase space is split into two parts, using the matrix
element cross sections in one region and the PS in the other
\cite{sey1,sey2,lepto}, or the PS algorithm is modified as to
reproduce the matrix element cross section in the hard limit \cite{bs}. 
In current event generators such as e.g.\ {\tt PYTHIA} \cite{pythia}
or {\tt HERWIG} \cite{herwig}, leading log parton showers are combined
with leading order matrix elements in the wide angle scattering region.
Recently attempts where made to improve the leading order accuracy in
these approaches by going to next-to-leading order. In \cite{mrenna}
both, the PS and the matrix elements were modified to obtain a smooth
merge of the PS to the higher order calculation. Collins \cite{2}
suggested a procedure to subtract from the matrix elements those parts
which are included already in the PS. Since the PS
describes the soft and collinear region, the divergences associated
with the matrix element calculation in this region are
avoided. Finally, Sj\"ostrand and Friberg \cite{1} suggested an
improvement of the splitting procedure, by using the NLO matrix
elements to calculate the weight for the PS region instead of the LO 
weight. They suggested to introduce a function which approximates the
weight in the PS region in such a way that the negative weight problem
is largely avoided. However, Sj\"ostrand and Friberg only gave an
outline of their method without providing the approximate function for
any specific process, and hence they also did not give numerical results.

In this paper, we take the suggestions in \cite{1} as a starting point
and give further details for a general method to get the correct NLO
normalization of the PS region. To avoid negative weights, we rely on
an older suggestion by Baer and Reno \cite{oldpc}, based on the
phase-space-slicing method. Baer and Reno suggested to introduce a
phenomenologically determined fixed slicing parameter such that the
sum of the Born, virtual, soft and collinear contributions is
approximately zero. In this paper, we {\it calculate} a  
scale dependent cut-off function, which provides a cut-off for each
phase-space point, such that the sum is {\it exactly} zero and
the improved scale and scheme dependence of the NLO calculations is
preserved. The NLO corrections are given by the NLO hard tree-level
matrix elements integrated down to the cut-off. Since these
contributions are positive definite, the NLO weight obtained from this
method can be directly redistributed according to a PS algorithm. 

The paper consists of two main parts. In the first part we
discuss the general method to combine matrix elements at NLO with PS's
by use of the described cut-off function. In the second part we
consider as a physically important and interesting example single-jet
production in deep-inelastic scattering (DIS). Here, the LO
contribution is of order $\alpha_s^0$ and the NLO corrections are of
order $\alpha_s$. Including higher order matrix elements for this
process will become especially important for diffractive DIS, since
the gluon density in the proton at small $x_{bj}$ is large compared to
the quark density. Therefore, the photon-gluon fusion process, which
is a NLO correction to the first order quark-parton model
contribution, will not necessarily be small. We explicitly construct
and numerically study the cut-off function for this case. Finally, we
summarize our results and give an outlook for future developments.

\section{General method}

\subsection{Jet cross sections at next-to-leading order}

We start by summarizing the procedure for the numerical evaluation of
an inclusive $n$-jet cross section in NLO QCD. The first step is to
select a jet algorithm, which defines how partons are recombined to
give jets. In the following we take for definiteness the invariant 
mass $s_{ij}$ of two partons $i$ and $j$ and define the $n$-jet
region such that $s_{ij}<\smin$, with some kind of minimum mass
$\smin$ and likewise the $(n+1)$-jet region such that $s_{ij}>\smin$ for
all $i,j$. The LO process for the production of $n$ jets consists of
$n$ final state partons and obviously does not depend on the jet
definition. This dependence only comes in at NLO. The ${\cal O}(\alpha_s)$ 
corrections to this process are given by the ultraviolet (UV) and
infrared (IR) divergent one-loop contributions to the $n$-parton
configuration, which are the virtual corrections, and the NLO tree
level matrix elements with $(n+1)$ partons, the real corrections. The
tree-level matrix elements have to be integrated over the phase space
of the additional parton, which gives rise to collinear and soft
singularities. After renormalization, the singularities in the virtual
and soft/collinear contributions cancel and remaining poles are
absorbed into parton distribution functions. One wants to integrate
most of the phase space of the real corrections numerically, but one
needs to find a procedure to calculate the soft/collinear
contributions analytically as to explicitly cancel the poles from the
virtual corrections. The two basic methods to perform these
integrations are the subtraction method  \cite{ERT,KS,CS} and the
phase-space slicing (PSS) method \cite{pps2,pps3,pps1,Gr} (see also
\cite{kramer} for a review).

In the following we will make use of the PSS method and therefore
discuss this method further. To illustrate the method, we rely on the
classical example given by Kunszt and Soper \cite{KS}. We label the LO
Born contribution as $\sigma^{\mboxsc{LO}}=\sigma^B$. The NLO cross
section is given by the sum of the Born cross section and the virtual
and real corrections, $\sigma^V$ and $\sigma^R$:
\bq
  \sigma^{\mboxsc{NLO}}=\sigma^B + \sigma^V + \sigma^R 
   = \sigma^B + C_V - \lim_{\epsilon\to 0} \frac{1}{\epsilon}F(0) +
   \int_0^1 \frac{dx}{x}F(x) \ . \label{KS-ex}
\eq 
Here, $F(x)$ is the known, but complicated function representing the
$(n+1)$-parton matrix elements. The variable $x$ represents an angle
between two partons or the energy of a gluon, the integral represents
the phase-space intergation that has to be performed over the
additional parton. The singularity of the real corrections at $x\to 0$ is
compensated by the virtual corrections, given by the pole term and some
constant, $C_V$. In the PSS method, the integral over the real
corrections is divided into two parts, $0<x <\delta$ and 
$\delta <x<1$. We note that the technical cut-off $\delta$ should
lie within the $n$ jet region, i.e., if we define
$\ycut=\smin/Q^2$, then we should have $\delta<\ycut$. If the
cut-off parameter is sufficiently small, $\delta \ll \ycut <1$, one
can write 
\begin{eqnarray}
  \sigma^R &=& \int_0^1 \frac{dx}{x}F(x) \simeq 
 \lim_{\epsilon \to 0} \left\{
 \int^1_{\delta} \frac{dx}{x} x^\epsilon F(x)
 +F(0) \int^{\delta}_0 \frac{dx}{x} x^\epsilon \right\} \nonumber \\ 
  &\simeq & \int^1_{\delta} \frac{dx}{x} F(x) +F(0) \ln(\delta) +
  \lim_{\epsilon\to 0} \frac{1}{\epsilon}F(0) \ ,
\end{eqnarray}
where the integral has been regularized by the term $x^\epsilon$, as
suggested by dimensional regularization. The pole is now explicit and
the NLO cross section $\sigma^{\mboxsc{NLO}}$ is finite:
\bq
  \sigma^{\mboxsc{NLO}} \simeq \sigma^B + C_V + \int_{\delta}^1
  \frac{dx}{x} F(x) + F(0)\ln(\delta) \ .  \label{ex1}
\eq

Clearly, the real corrections $\sigma^R$ should not depend on $\delta$,
and the logarithmic $\delta$ dependence of the last term in eqn
(\ref{ex1}) should be canceled by the integral, which sometimes is
numerically difficult for very small parameters $\delta$. However, an
improvement of the above solution is possible by using a hybrid of the
PSS and the subtraction methods, suggested by Glover and Sutton
\cite{GSu}. In this method, one adds and subtracts only the universal
soft/collinear approximations for $x < \delta$, such that 
\begin{eqnarray}
 \sigma^R &= & \lim_{\epsilon \to 0} \left \lbrace
 \int^1_0 \frac{dx}{x} x^\epsilon F(x)
  -F(0) \int^{\delta}_0 \frac{dx}{x} x^\epsilon
  +F(0) \int^{\delta}_0 \frac{dx}{x} x^\epsilon
  \right \rbrace \nonumber \\
    &\simeq & \int^1_{\delta} \frac{dx}{x} F(x) 
   + \int^{\delta}_0 \frac{dx}{x} \bigg[ F(x)-F(0) \bigg] 
   + F(0) \ln(\delta) +
     \lim_{\epsilon\to 0} \frac{1}{\epsilon}F(0)  \ .
\end{eqnarray}
A cancellation between the analytical and numerical terms still
occurs, however only the phase space is approximated, so that this
method is valid at larger values of $\delta$. In the case where the
phase-space is not approximated for small $x$ the hybrid method
becomes independent from $\delta$.

\subsection{Event generation with positive NLO weights}

We would now like to improve the above NLO jet cross section by 
including the PS in the $n$-jet region. We keep the jet definition,
which separates the available phase space into two complementary
regions, namely the $n$-jet region for $s_{ij}<\smin$ and the
$(n+1)$-jet (hard) region for $s_{ij}>\smin$. The  NLO corrections,
i.e., the virtual and the soft/collinear corrections, will occur in
the $n$-jet region, so that the $n$-jet exclusive cross 
section has a NLO normalization. On the other hand, the hard $(n+1)$-jet
region contains tree-level contributions only.

In current event generators, a weight will be generated in the
$n$-jet region by calculating the LO, tree-leel matrix element and
then redistributing this weight with help of the PS algorithm. The
same jet criterion used to separate the matrix element and PS regions is
used to veto events from the PS which would lie outside the $n$-jet
region, as to avoid double counting. The $(n+1)$-jet region is described
by the hard $(n+1)$-parton matrix elements. 

If we take the procedure for calculating the NLO corrections as
described in the previous section, we could in principle calculate the
weight in the $n$ jet region in NLO. However, we would like to avoid
the generation of large positive and negative weights, since this is
not a practical starting point for an event generator, especially if
one wishes to add hadronization after the showering. Therefore we
suggest to choose the PSS parameter $\delta$ such that the weights
are always positive. To keep most of the advantages
of the NLO calculation, we aim to find a cut-off function 
$\deltan (\mu^2)$ that provides a cut-off parameter for any given
renormalization and factorization scale which lies inside the $n$ jet
region,  such that the sum of the Born, soft/collinear and virtual
contributions are exactly zero. The NLO corrections inside the jet are
then completely enclosed in the $(n+1)$-parton hard matrix elements,
integrated down to the cut-off function. Only the tree-level matrix
elements will serve as starting points for the PS. The reduced scale
and scheme dependence of the NLO cross sections is subsumed into the
scale dependent cut-off function. 
Staying in the simple example from eqn (\ref{KS-ex}), the aimed
cut-off function reads
\bq
   \deltan = \exp\left( -\frac{\sigma^B}{F(0)}- \frac{C_V}{F(0)} \right) \ .
   \label{sol}
\eq 
Substituting $\delta$ in eqn (\ref{ex1}) by $\deltan$ results in 
\bq
  \sigma^{\mboxsc{NLO}} \simeq \int\limits_{\deltan}^1 \frac{dx}{x} F(x) \ ,
 \label{h1}
\eq 
which is exactly what we are looking for. The NLO corrections
are given completely by the hard $(n+1)$-parton tree-level matrix elements,
which are positive definite and can be combined with the PS in the usual
way.  The function $\deltan$ will depend on the kinematics of the $n$
parton configuration, as  well as on the renormalization and
factorization scales. Of course, one has to ensure in the event
generation process that $\deltan<\ycut$, i.e.\ that the cut-off lies
within the $n$-jet region. 

In general, the function (\ref{sol}) could yield a relatively large
$\delta$, so that the simple PSS method fails to give the correct
answer for the NLO cross section.\footnote{As we will see in the
second part of the paper, for DIS single-jet production at ${\cal
O}(\alpha_s)$ the cut-off $\deltan$ actually is sufficiently small for
the soft and collinear approximations to be valid.} We can however
calculate the missing parts of the approximated matrix elements, which
is the cause of the error, numerically by use of the hybrid method,
described above. Improving the result (\ref{h1}) with the hybrid
method, one finds 
\bq
   \sigma^{\mboxsc{NLO}} \simeq \int\limits_{\deltan}^1 \frac{dx}{x} F(x) 
   + \int\limits^{\deltan}_0 \frac{dx}{x} \bigg[ F(x)-F(0) \bigg] \ ,
  \label{ex2}
\eq 
which gives additional terms of order $\deltan\ln(\deltan)$. Here, the
difference between the approximate function $F(x)|_{x=0}$ and the full
expression is evaluated numerically. This is similar to the what
Sj\"ostrand and Friberg suggested in \cite{1} as a starting point for
the event generation.

To summarize, we propose in close analogy to \cite{1} the following
steps to produce an event in the PS region with positive weight and
NLO normalization:
\begin{itemize} 
\item Define the $n$-jet region by 
 \bq  s_{ij} < \ycut Q^2 \eq
  for all partons $i,j$. The $(n+1)$-jet region given by
  $s_{ij}>\ycut$ is described by the hard $(n+1)$-parton matrix elements.
\item For each $n$-jet phase-space point calculate $\deltan$ according
  to eqn (\ref{sol}). The cross section for the $n$-jet region is then
  given by  
  \bq \sigma^{NLO}_{n-jet} = \sigma^B+\sigma^V+\sigma^R 
       = \int\limits_{\deltan}^{\ycut} \frac{dx}{x} F(x)  
   + \int\limits^{\deltan}_0 \frac{dx}{x} \bigg[ F(x)-F(0) \bigg] 
  \eq
  We can write this symbolically as 
  \bq \sigma^{NLO}_{n-jet} 
       = (\ycut-\deltan) \left<\frac{F(x)}{x}\right>
   + \deltan \left< \frac{F(x)-F(0)}{x} \right>
  \eq
  as to express the Monte-Carlo nature of the procedure.
\item In the $n$-jet region generate an $x_k=s^k_{ij}/Q^2$ with 
  $x_k\in [0,\ycut]$ (the index $k$ refers to the $k$th event). The
  weight which will be redistributed by the parton shower algorithm is
  given by
  \[ W_k =(\ycut-\deltan) \left<\frac{F(x)}{x}\right>\bigg|_{x=x_k} \]
  for $x_k\in [\deltan,\ycut]$, where $W>0$ by construction, or by
  \[ W_k = \deltan \left< \frac{F(x)-F(0)}{x} \right>\bigg|_{x=x_k} \]
  for $x_k\in [0,\deltan]$, where $W$ will be rather small.
\item Use the same jet resolution criterion as in the first step to
veto events from the PS that lie within the $(n+1)$-jet region.
\end{itemize}

\section{Inclusive single-jet production in DIS}

In the following we apply the ideas of the previous section to inclusive
single-jet production in DIS $eP$-scattering at ${\cal
O}(\alpha_s)$. For this, we recall the formul\ae\ for calculating
single-jet cross sections at NLO with the standard PSS method and
describe the modified version of the PSS method, which is used to
evaluate the cut-off function that can easily be used in  existing
event generators for electroproduction. We numerically compare the
modified with the standard method.

\subsection{Single-jet cross section up to ${\cal O}(\alpha_s)$}

In $eP$-scattering
\bq
  e(k) + P(p) \to e(k^\prime) + X
\eq
the final state with a single jet is the most basic event with a large
transverse energy $E_T$ in the laboratory frame. The lowest order
$\O(\alpha_s^0)$ partonic contribution to the single-jet cross section
arises from the quark parton model (QPM) subprocess
\bq
 e(k)+q(p_0)\rightarrow e(k^\prime) + q(p_1)
\label{qtoq}
\eq
and the corresponding anti-quark process with $q\leftrightarrow \bar{q}$.
The partonic cross section for this process is given by 
\bq
\hat{\sigma}^{\mboxsc{LO}}_{q\rightarrow q}
=\sigma_0\,\,e_q^2\,\,|M_{q\rightarrow q}|^2
\label{siglodef}
\eq
where
\bq
 \sigma_0 =\frac{1}{4p_0.k}\, \frac{1}{4}\,\frac{(4\pi\alpha)^2}{Q^4}
 \label{sigma0def}
\eq
and
\begin{equation}
|M_{q\rightarrow q}|^2=32\,
\left[(p_0.k)^2 \, +\,(p_0.k^\prime)^2 \,\right]
\,= \,8 \hat{s}^2\,\,(1+(1-y)^2)
\label{m_qtoq}
\end{equation}
Here, $\hat{s}=x s$, with $s = (k+p)^2$, denotes the partonic center of
mass energy squared, $\alpha$ is the electromagnetic coupling and
$Q^2=2(k.k^\prime)$ is the photon virtuality. The total DIS cross
section can be directly obtained by integrating out the complete phase
space of the final state parton.

At NLO, the single-jet cross section receives contributions from the
real and the one-loop virtual corrections. The real corrections
consist of the 
photon-gluon fusion and the QCD-Compton processes 
\begin{eqnarray}
e(k)+q(p_0)&\rightarrow& e(k^\prime) + q(p_1) + g(p_2)
\label{qtoqg} \ ,\\
e(k)+g(p_0)&\rightarrow& e(k^\prime) + q(p_1) + \bar{q}(p_2) \ ,
\label{gtoqqbar}
\end{eqnarray}
together with corresponding anti-quark processes. The partonic cross
sections for one-photon exchange are
\begin{eqnarray}
\alpha_s\, \hat{\sigma}^{\mboxsc{LO}}_{q\rightarrow qg}\,\, &=&
\sigma_0\,\,e_q^2\,\,
(4\pi\alpha_s(\mu_R))\,\,|M_{q\rightarrow qg}|^2 \ ,
\label{sig_qtoqg} \\[1mm]
\alpha_s \, \hat{\sigma}^{\mboxsc{LO}}_{g\rightarrow q\bar{q}}\,\, &=&
\sigma_0\,\,e_q^2\,\,
(4\pi\alpha_s(\mu_R))\,\,
|M_{g\rightarrow q\bar{q}}|^2 \ ,
\label{sig_gtoqqbar}
\end{eqnarray}
with
\begin{eqnarray}
 |M_{q\rightarrow qg}|^2 &=&
 \frac{128}{3}\,\,(k.k^\prime)\,\,
 \frac{(k.p_0)^2+(k^\prime.p_0)^2+(k.p_1)^2+(k^\prime.p_1)^2
     }{(p_1.p_2)(p_0.p_2)}  \label{m_qtoqg} \ , \\[1mm]
 |M_{g\rightarrow q\bar{q}}|^2 &=& -\frac{3}{8}\,\,
 |M_{q\rightarrow qg}|^2(p_0\leftrightarrow-p_2)\nonumber\\ &=&
 16\,\,(k.k^\prime)\,\,
 \frac{(k.p_2)^2+(k^\prime.p_2)^2+(k.p_1)^2+(k^\prime.p_1)^2}
      {(p_0.p_1)(p_0.p_2)} \ .
\label{m_gtoqqbar}
\end{eqnarray}
Color factors (including the initial state color average) 
are included in the squared matrix elements. Note that the initial
state spin average factors are included in the definition of
$\sigma_0$ in eqn~(\ref{m_qtoq}) and that the results in
eqns~(\ref{m_qtoqg},\ref{m_gtoqqbar}) contain  the full polarization
dependence of the virtual boson.

As already discussed in the previous section, the real corrections
inherit characteristic divergencies, which are the initial and final
state soft and collinear singularities. These  can be separated from
the hard phase-space regions by introducing a 
cut-off parameter $\smin$ \cite{pps2,pps3,pps1,Gr,kramer}. The hard
part can be integrated numerically, whereas the soft/collinear part is
treated analytically. The analytical integrals can be performed in
$n=4-2\epsilon$ dimensions. The poles which appear in $\epsilon$
cancel against poles from the one-loop corrections. Remaining poles in
the initial state are proportional to the Altarelli-Parisi splitting
functions and are absorbed into the parton distribution functions
(PDF's) of the proton, $f_i(x,\mu_F)$ for $i=q,\bar{q},g$. UV
divergencies in the one-loop corrections are absorbed into the running
coupling constant $\alpha_s(\mu_R)$.

The ${\cal O}(\alpha_s)$ corrections to the ${\cal O}(\alpha_s^0)$ Born term 
are known for quite some time \cite{fxnlo0} and the one-jet inclusive
final states have been discussed in \cite{Gr,nlo0}. Since we will
later on rely on programs provided together with the {\tt MEPJET}
Monte Carlo \cite{mepjet} for tabulating the integrals that occur for the 
initial state corrections, we will here apply the method of crossing
functions as used in \cite{mepjet} and outline in \cite{mirkes}, which
is fully equivalent to the results in \cite{Gr,nlo0}. We
take over the notation in \cite{mirkes}. The finite part of 
the NLO partonic cross section, which is arrived at by summing up the
virtual contributions and the singular parts of the two-parton final
state is given by the expression
\bq
 \alpha_s \,\hat{\sigma}^{\mboxsc{NLO}}_{q\rightarrow q}\,\, =
 \alpha_s \,\,\sigma_0\,\,e_q^2\,\,
 |M_{q\rightarrow q}|^2\,\,
 {\cal{K}}_{q\rightarrow q}(\smin,Q^2) \ . \label{nlo1j}
\eq
The finite parts of the virtual corrections factorize the Born matrix
element. The factor ${\cal{K}}_{q\rightarrow q}$, depending on both
$\smin$ and the invariant mass of the hard  partons  $2p_0.p_1=Q^2$,
is given by 
\begin{equation}
 {\cal{K}}_{q\rightarrow q}(\smin,Q^2) = \frac{8}{9}\,
 \left(\frac{N_C}{2\pi}\right)
 \left[
 -\ln^2\left( \frac{\smin}{Q^2}\right) 
 - \frac{3}{2}\ln\left(\frac{\smin}{Q^2}\right)
 -\frac{\pi^2 }{3}-\frac{1}{2}
 \, + \, {\cal{O}}(\smin) \right]  \ , \label{r_qtoq}
\end{equation}
where $N_C=3$ is the number of colours. ${\cal{K}}_{q\rightarrow q}$ may be
crossed in exactly the same manner as the usual tree level crossing 
from the ${\cal{K}}$ factor in $e^+e^-\rightarrow$ 2 partons as
given in eqn~(4.31) with $n=0$ in Ref~\cite{pps1} or in eqn~(3.1.68)
of \cite{kramer}. Thus, eqn~(\ref{r_qtoq}) includes also the
crossing of a pair of collinear partons with an invariant mass smaller
than $\smin$ from the final state to the initial state.  This
`wrong' contribution is replaced by the correct collinear initial
state configuration by adding the appropriate crossing function
contribution to the hadronic cross section, which takes also into
account the corresponding factorization of the initial state
singularities, encoded in the crossing functions
$C_q^{\overline{\mboxsc{MS}}}$ for valence and sea quark 
distributions. The  crossing functions for an initial state parton
$a$, which participates in the hard scattering process, can be
written in the form \cite{mirkes}
\bq
C_{a}^{\overline{\mboxsc{MS}}}(x,\mu_F,\smin)=
\left(\frac{N_C}{2\pi}\right)
\left[ A_{a}(x,\mu_F)\ln\left(\frac{\smin}{\mu_F^2}\right)
+      B_{a}^{\overline{\mboxsc{MS}}}(x,\mu_F)\right] \ , 
\label{crossf}
\eq
with
\bq
A_a(x,\mu_F) = \sum_p A_{p\rightarrow a}(x,\mu_F)
\eq
and
\bq
B_a^{{\overline{\mboxsc{MS}}}}(x,\mu_F) = \sum_p
B_{p\rightarrow a}^{{\overline{\mboxsc{MS}}}}(x,\mu_F) \ . 
\eq
The sum runs over $p=q,\bar{q},g$. The individual functions $A_{p\to
a}(x,\mu_F)$ and $B_{p\to a}^{{\overline{\mboxsc{MS}}}}(x,\mu_F)$ are
stated in the appendix.  In particular  all plus prescriptions
associated with the factorization of the initial state collinear
divergencies are absorbed in the crossing functions
$C_q^{\overline{\mboxsc{MS}}}$ which  is very useful for a Monte Carlo 
approach. We note that although the two-parton final state
contributions (\ref{m_qtoqg}) and (\ref{m_gtoqqbar}) contain the
full polarization dependence of the virtual photon, the singular
contributions occur only for the transverse photon polarization. 

Taking into account now virtual, initial and final state corrections
we can write the hadronic cross section for the one-parton final state
up to $\O(\alpha_s)$ as
\begin{eqnarray}
 \sigma^{\mboxsc{1parton}}_{\mboxsc{had}}(\smin) 
 &=& \sigma_0 \
 \sum_{i=q,\bar{q}} e_i^2\ \int dx \ d{\mbox{PS}}^{(k^\prime+1)} \ 
 \bigg[ f_i(x,\mu_F) \left(1 + \alpha_s(\mu_R)\ {\cal{K}}_{q\rightarrow
 q}(\smin,Q^2) \right) \nonumber \\
 &+& \alpha_s(\mu_R)\ C_i^{\overline{\mboxsc{MS}}}(x,\mu_F,\smin) \bigg]
 |M_{q\rightarrow q}|^2  \label{1parton} \ .
\end{eqnarray}
To obtain the final, $\smin$ independent result, one also has to add the
contribution containing the two parton final state, integrated over
those phase-space regions, where any pair of partons $i,j$ with
$s_{ij}=(p_i+p_j)^2$ has $s_{ij}>\smin$:
\begin{eqnarray}
 \sigma^{\mboxsc{2parton}}_{\mboxsc{had}}(\smin) &=& \sigma_0 \
 \sum_{i=q,\bar{q}} e_i^2 \!
 \int\limits_{|s_{ij}|>\smin} \!\!\!\!\!\!  dx \ d\mbox{PS}^{(k^\prime+2)}
 \ 4\pi\alpha_s(\mu_R)\ \bigg[ f_i(x,\mu_F)] \ |M_{q\rightarrow qg}|^2
 \nonumber \\ &+&  \mbox{$\frac{1}{2}$} \, f_g(x,\mu_F) \ 
 |M_{g\rightarrow q\bar{q}}|^2 \bigg] \ .  \label{2parton}
\end{eqnarray}
The Lorentz-invariant phase space measure
$d\mbox{PS}^{(k^\prime+n)}$ contains both the scattered electron and
the partons from the photon-parton scattering process and is defined
as 
\bq
 d\mbox{PS}^{(k^\prime+n)}=\delta^4(p_0+k-k^\prime-\sum_{i=1}^{n}p_i)
 \,2\pi\frac{d^3k^\prime}{2E^\prime}
 \prod_{i=1}^{n}\frac{d^3 p_i}{(2\pi)^3\,2E_i} \ .
\label{phasespace}
\eq
The bremsstahlung contribution in (\ref{2parton}) grows with
$\ln^2\smin$ and $\ln\smin$ with decreasing $\smin$. Once $\smin$ is
small enough for the soft and collinear approximations to be valid, 
this logarithmic growth is exactly canceled by the explicit
$-\ln^2\smin$ and $-\ln\smin$ terms in ${\cal{K}}_{q\rightarrow q}$
and the $\smin$ dependence in the crossing functions.

\subsection{Cut-off function}

We are now in the position to reformulate the PSS
method for the single-jet inclusive cross section for our purposes. As
explained in section 2, we wish to avoid the NLO one-parton
contributions contained in eqn~(\ref{1parton}) completely. Integrating
out the delta-function (\ref{phasespace}) in eqn (\ref{1parton}) we
obtain, omitting scale dependences, 
\bq
 \frac{d\sigma^{\mboxsc{1parton}}_{\mboxsc{had}}}{dx\,dQ^2} = 
 \frac{2\pi\alpha}{xQ^4} \ (1+(1-y)^2) \ \sum_{i=q,\bar{q}} e_i^2 \
 x \left[ f_i(x) \left( 1+\alpha_s\ {\cal{K}}_{q\to q}(Q^2)\right) + \alpha_s\
 C_i^{\overline{\mboxsc{MS}}}(x) \right] \ .
 \label{hlp}
\eq
The $\smin$-dependence of this one-parton cross section is
canceled by the respective unresolved two-parton cross section for each
phase-space point $(x,Q^2)$. In order to avoid the one-parton final states,
it will be sufficient to chose an appropriate value
of the cut-off parameter (which we denote as $\sminn$) for each
phase-space point $(x,Q^2)$, so that
\bq
  \frac{d\sigma^{\mboxsc{1parton}}_{\mboxsc{had}}}{dx\,dQ^2}(\sminn) = 0
  \ .  \label{start}
\eq
To solve eqn~(\ref{start}) for $\smin$, it is sufficient to solve the
equation
\bq
 \sum_{i=q,\bar{q}} e_i^2\ \bigg[ f_i(x,\mu_F)
 \left(1 + \alpha_s(\mu_R)\ {\cal{K}}_{q\rightarrow  q}(\smin,Q^2) \right)
 + \alpha_s(\mu_R)\ C_i^{\overline{\mboxsc{MS}}}(x,\mu_F,\smin) \bigg] = 0
 \ .  \label{step}
\eq
The $\smin$ dependence of ${\cal{K}}_{q\rightarrow  q}$ can be seen in
eqn~(\ref{r_qtoq}), whereas the $\smin$ dependence of
$C_i^{\overline{\mboxsc{MS}}}$ is given in eqn~(\ref{crossf}). For
convenience, we define the sums
\begin{eqnarray}
 F &=& \sum_{i=q,\bar{q}} e_i^2\ f_i(x,\mu_F) \ , \\ 
 A &=& \sum_{i=q,\bar{q}} e_i^2\ A_i(x,\mu_F) \ , \\
 B &=& \sum_{i=q,\bar{q}} e_i^2\ B_i^{{\overline{\mboxsc{MS}}}}(x,\mu_F) 
\end{eqnarray}
and the functions
\begin{eqnarray}
  \eta &=& \ln\left(\frac{Q^2}{M^2}\right) - \frac34 + \frac{9}{16}
  \frac{A}{F} \ , \\ 
  \psi &=&  -\ln^2\left(\frac{Q^2}{M^2}\right) +
  \frac{3}{2}\ln\left(\frac{Q^2}{M^2}\right)
  -\frac{\pi^2}{3}-\frac{1}{2} + \frac98 \left[ \frac{2\pi}{N_C\alpha_s} +
  \frac{B}{F} - \frac{A}{F} \ln\left(\frac{\mu_F^2}{M^2}\right) \right] \ ,
\end{eqnarray}
which are independent of $\smin$ up to ${\cal{O}}(\smin)$. We have
introduced some arbitrary scale $M^2$ to keep the functions $\eta$ and
$\psi$ dimensionless. The solution of eqn~(\ref{step}) is then given
by the solution of the quadratic equation 
\bq
 \ln^2\left(\frac{\smin}{M^2}\right)  -  2\eta
 \ln\left(\frac{\smin}{M^2}\right) = \psi \ .  
\eq
We find for $\sminn$
\bq
 \sminn(\mu_F,\mu_R,x,Q^2) = \exp \left[ \ln(M^2) + \eta - \sqrt
  { \eta^2 + \psi } \,\, \right] \  , 
 \label{master}
\eq
where we have taken the smaller of the two solutions, since we require
$\smin$ to be sufficiently small for the soft and collinear
approximations to be valid. The $\ln(M^2)$ dependence in (\ref{master}) 
cancels in the sum of the individual terms in the exponent.

Inserting the $\sminn$ function into eqn~(\ref{2parton}) as a lower
integration boundary for each phase space point $(x,Q^2)$ will give
the complete answer for the single-jet cross section in NLO. This is
well suited for the purpose of combining matrix elements in NLO with
the PS.  It is important to note that the $\sminn$ function depends on
the factorization and renormalization scales, so that the improved
scale dependence of the NLO cross section is preserved in our modified 
approach. A crucial point, which we will study in detail in the next
section, is whether the $\sminn$ function obtained with
eqn~(\ref{master}) is small enough for the soft and 
collinear approximations, made to evaluate the expressions
(\ref{r_qtoq}) and (\ref{crossf}), to be valid.

\subsection{Numerical results}

In this section we numerically investigate the solution (\ref{master}).
We look at the size of $\sminn$ for given $x$ and $Q^2$ and study the
effect of scale changes on $\sminn$. Furthermore, we check whether
NLO single-jet inclusive cross sections obtained by integrating out
the two-parton contributions down to $\sminn$ gives the same result as
in the conventional approach, where one-parton and two-parton
contributions, separated by some fixed $\smin$, are summed. 

\begin{figure}
  \unitlength1mm
  \begin{picture}(122,190)
    \put(3,90){\epsfig{file=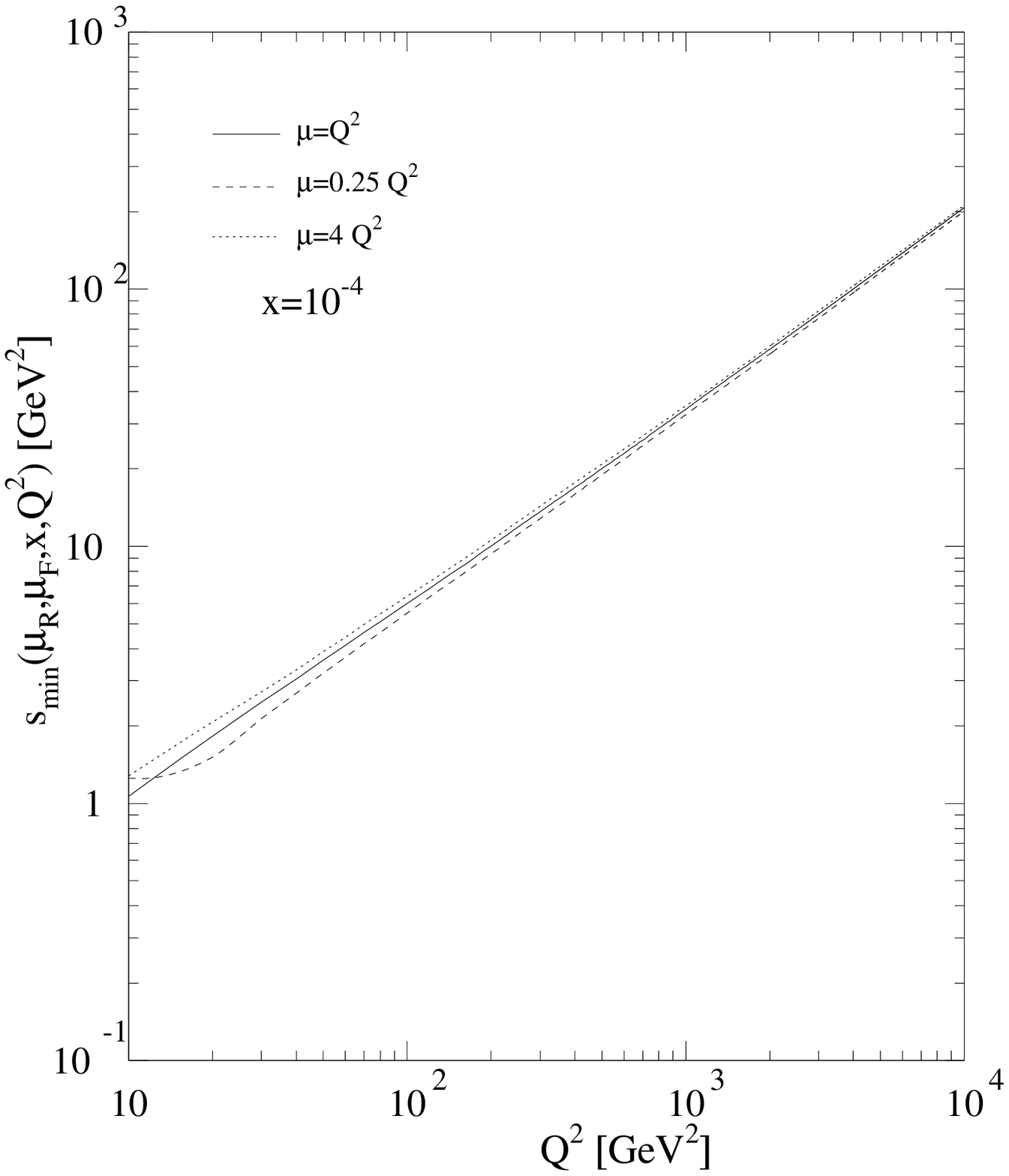,width=8cm}}
    \put(83,90){\epsfig{file=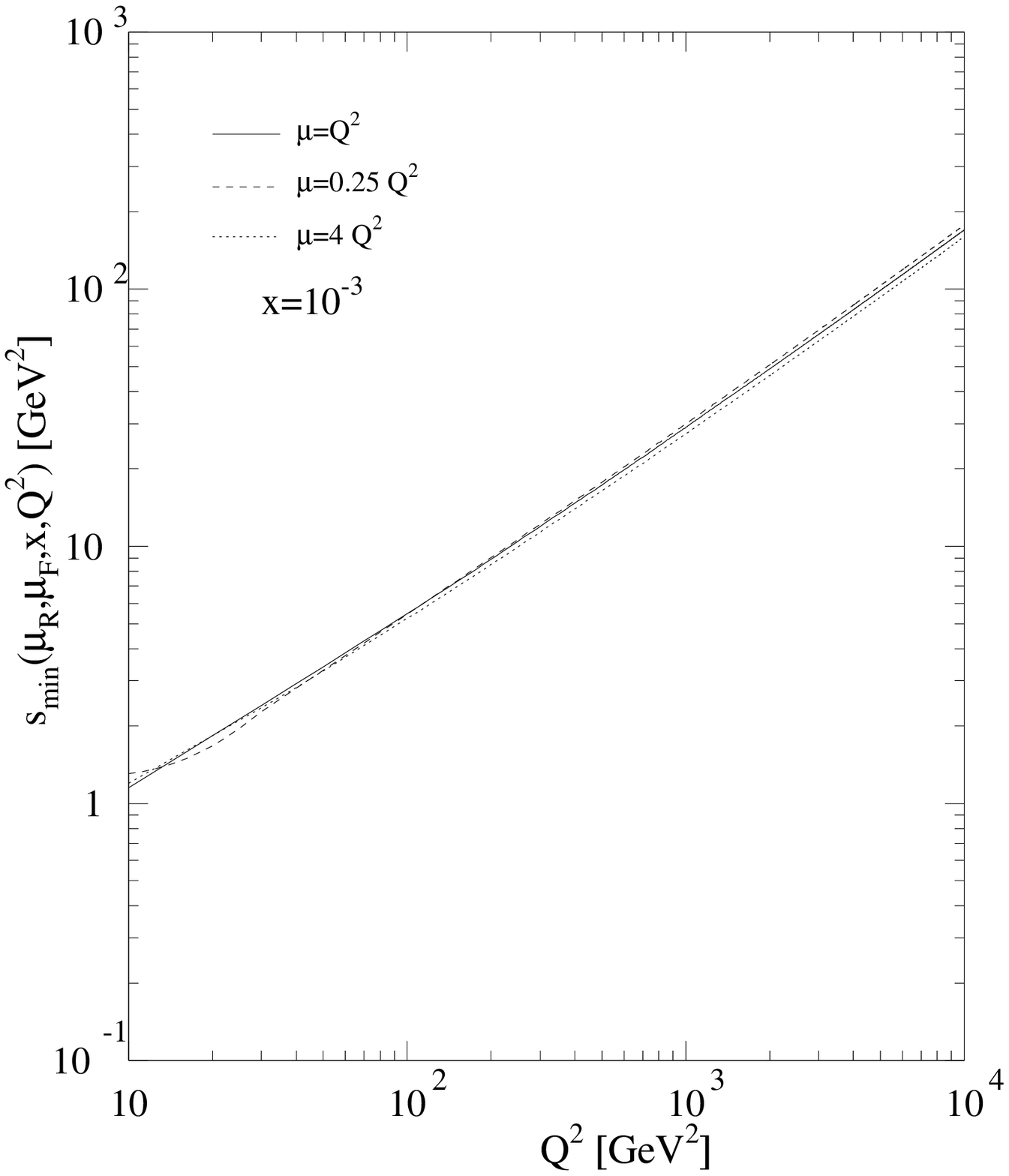,width=8cm}}
    \put(3,0){\epsfig{file=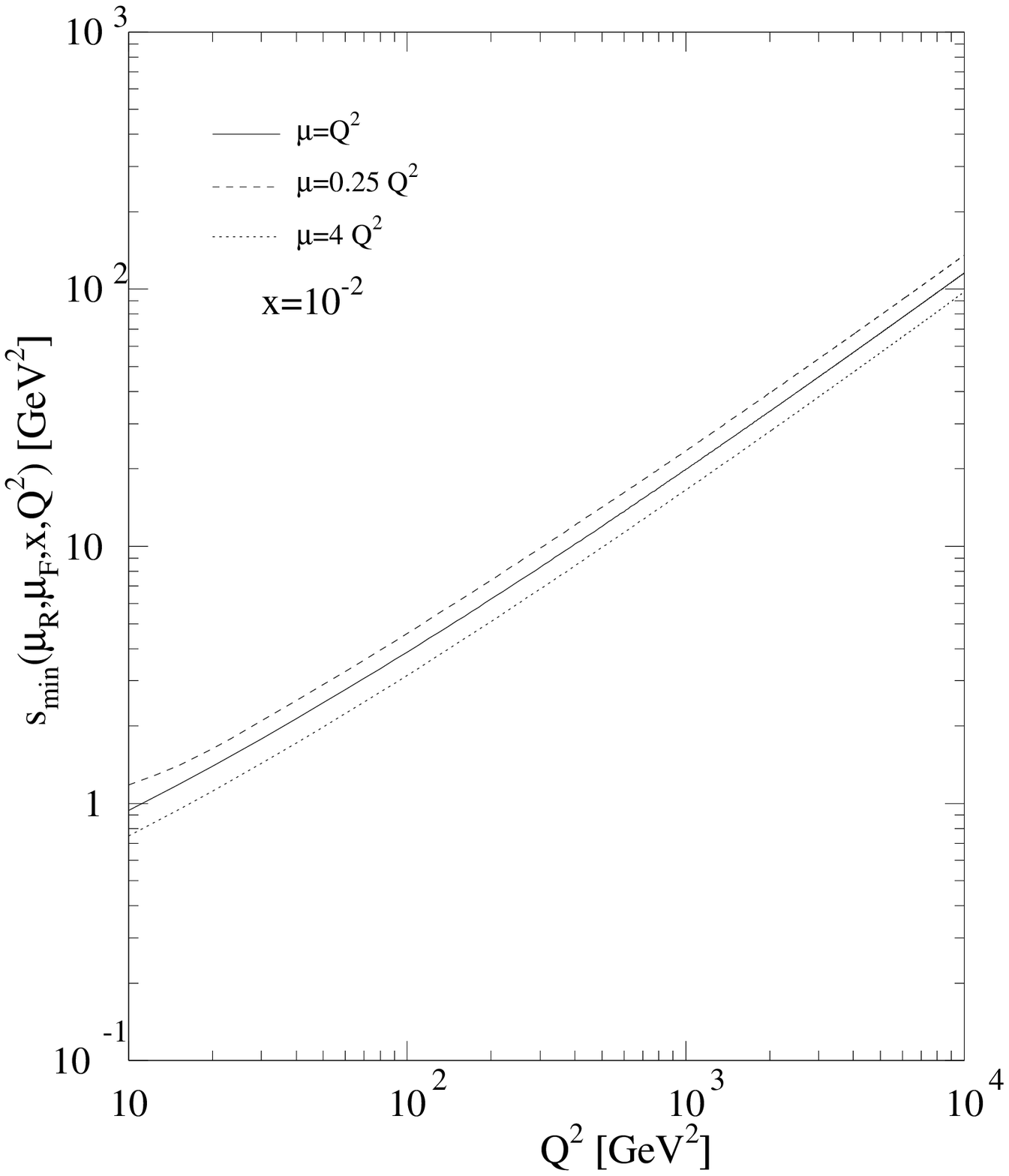,width=8cm}}
    \put(83,0){\epsfig{file=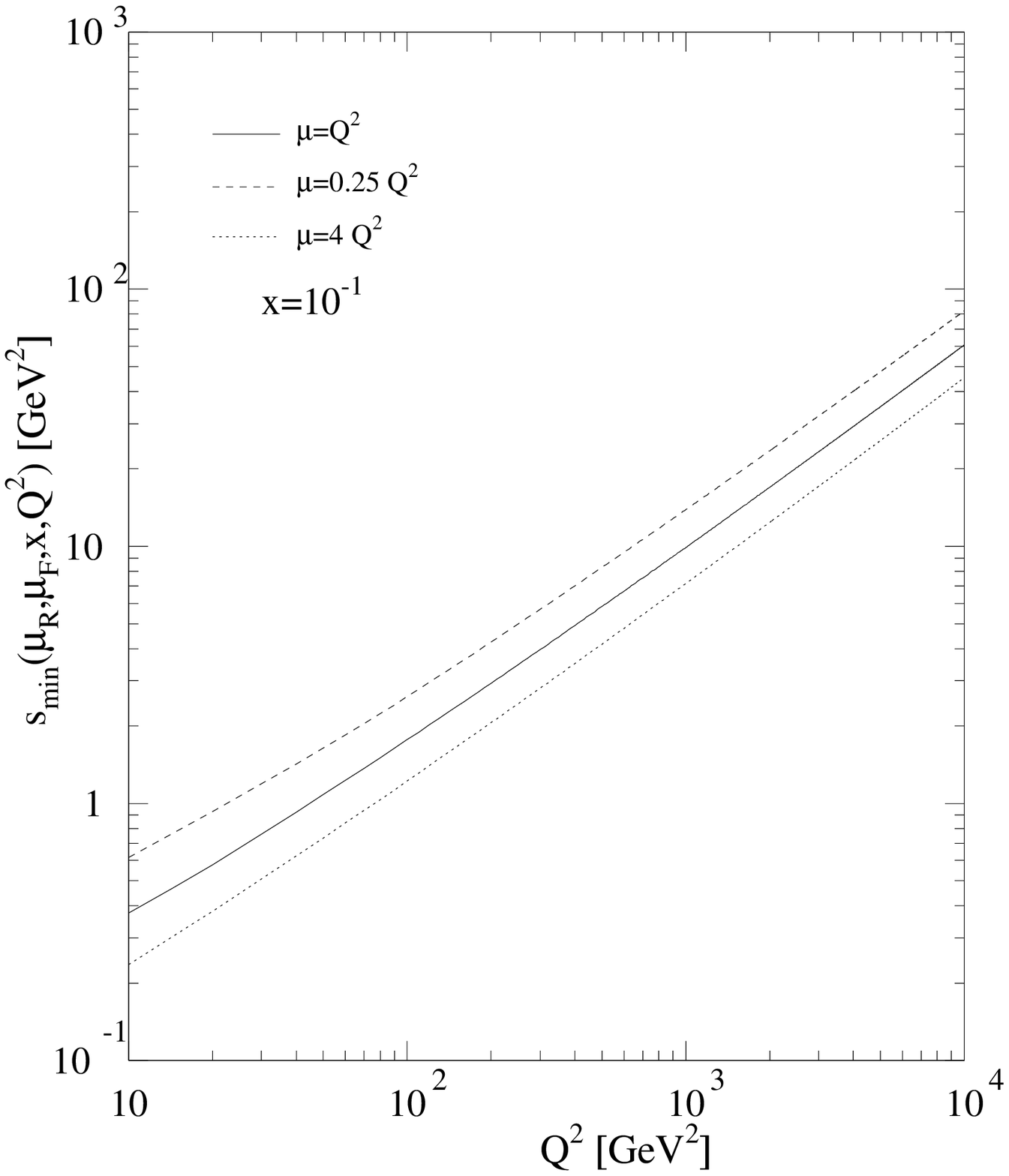,width=8cm}}
  \end{picture}
\caption{The function $\smin$ versus $Q^2$ for the scales
  $\mu_R=\mu_F=\mbox{$\frac14$}Q^2,Q^2,4Q^2$ for the four values
  $x=10^{-4},10^{-3},10^{-2}$ and $10^{-1}$.\label{f1}}
\end{figure}

We start by looking at the $\sminn$ function in the region given by 
$x\in [10^{-4},10^{-1}]$ and $Q^2\in [10,10^4]$~GeV$^2$. We produce
all results for one-photon exchange, i.e., neglecting possible
contributions from $Z$-exchange. We employ the 
MRST \cite{mrst} parton distributions for the proton and use the
integration package provided with {\tt MEPJET} to calculate and
tabulate the crossing functions \cite{mepjet} for these parton
distributions. This makes it numerically very convenient to use the
function $\sminn$, eqn (\ref{master}).\footnote{The FORTRAN code for the
$\sminn$ function can be obtained upon request from the author.}
In Fig.~\ref{f1} we have plotted $\sminn$ as a function of $Q^2$ for
the four fixed values $x=10^{-4},10^{-3},10^{-2}$ and $10^{-1}$ for
the scales $\mu=\mu_R=\mu_F=\xi Q^2$ with $\xi = \frac14,1$ and
$4$. We find values around $2$~GeV$^2$ in the small $Q^2$ region,
whereas they rise up to values between $100$ and $200$~GeV$^2$ for the
largest $Q^2$ values. The $\smin$ values are larger for smaller
$x$. The scale variation leads to small changes of the $\smin$
values. The scale variation in the actual cross sections will be still
smaller, since the $\smin$ dependence of the cross sections is
logarithmic. For the two larger $x$ values, the smaller scales leads
to a larger value of $\sminn$ which will therefore produce smaller
cross sections. For the two smaller $x$ values there seems to be a
compensation between the renormalization and factorization scale
variations, leading to a very small overall variation in $\sminn$,
especially at large $Q^2$.

Next, we numerically compare the standard PSS method
with our modified approach. The following comparisons are done for HERA
conditions, i.e., $E_e=27.5$~GeV and $E_p=820$~GeV, giving
$\sqrt{s}=300$~GeV. A cut of $E_e^\prime>10$~GeV is
applied to the final state electron and we choose $y\in [0.04,1]$. We
take the same $Q^2$ region as above, namely $Q^2\in [10,10^4]$~GeV$^2$.
Jets are defined in the laboratory frame with the $k_T$ algorithm with 
$E_T^{\mbox{\scriptsize lab}}>5$~GeV and $|\eta^{\mbox{\scriptsize lab}}|<2$.
All cuts together restrict the $x$ range to be $x\in [10^{-3},1]$. The
numerical results for the one-jet inclusive cross sections in the
following are produced with {\tt MEPJET}~\cite{mepjet}.

\begin{figure}
  \unitlength1mm
  \begin{picture}(122,190)
    \put(3,90){\epsfig{file=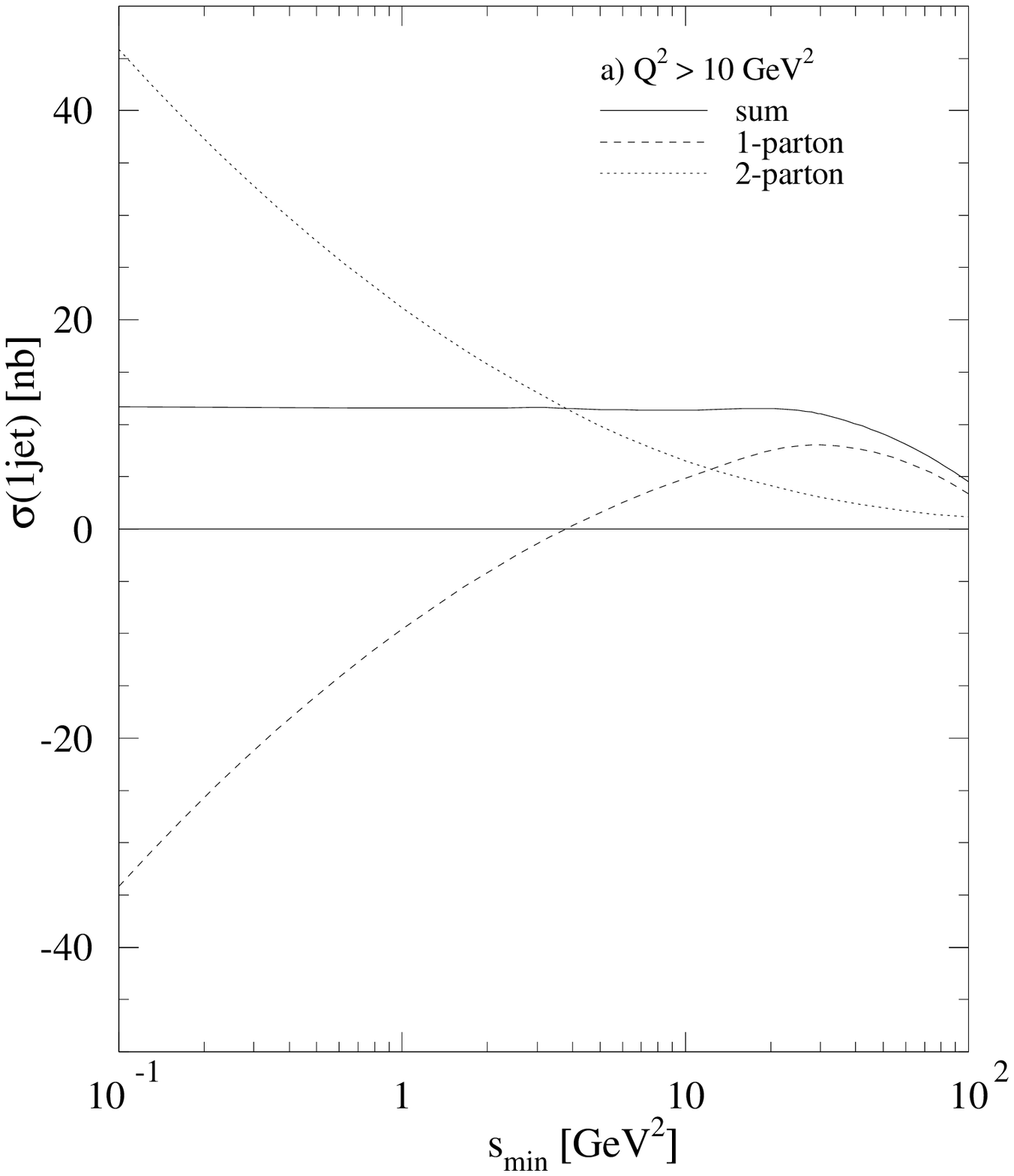,width=8cm}}
    \put(83,90){\epsfig{file=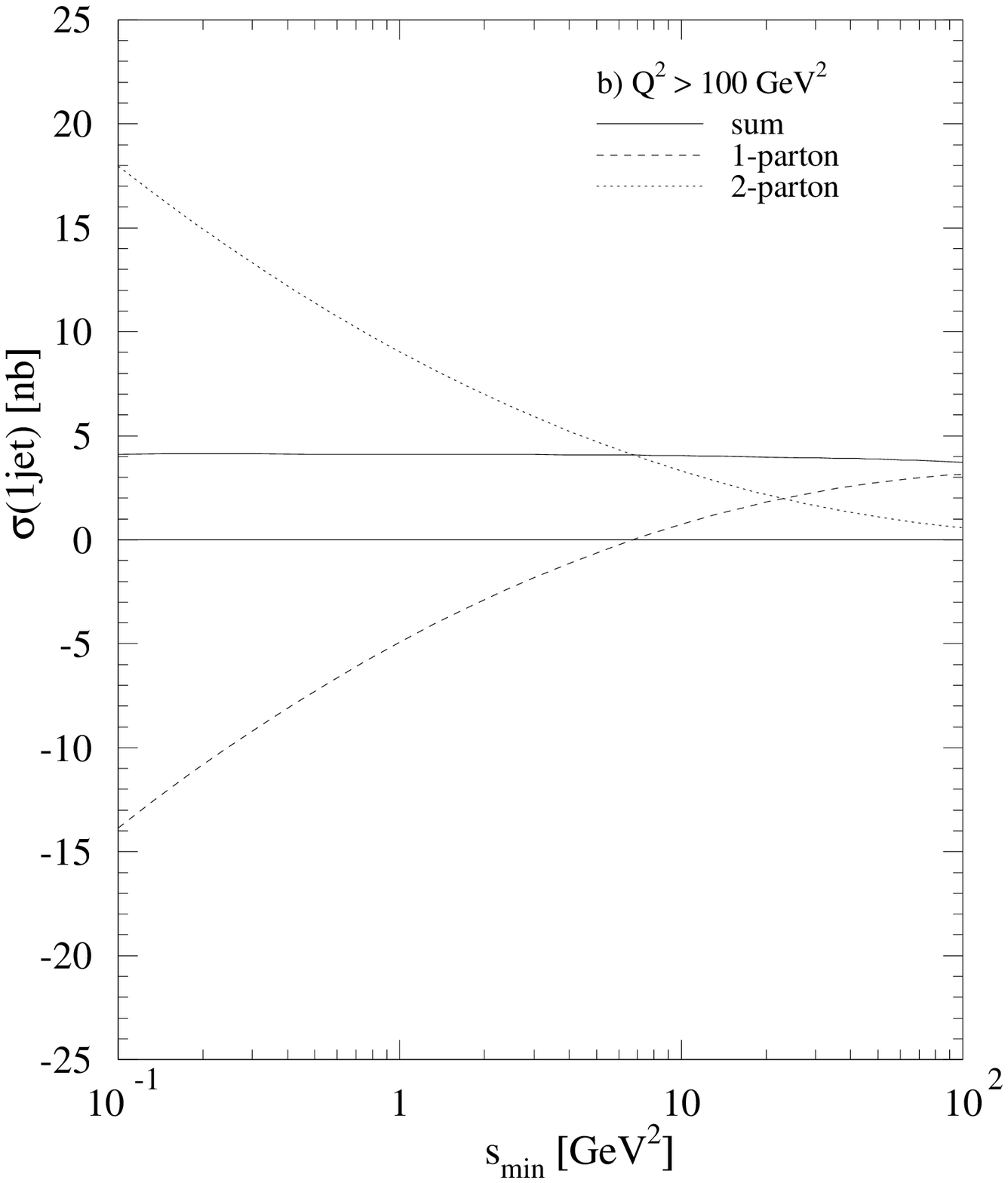,width=8cm}}
    \put(3,0){\epsfig{file=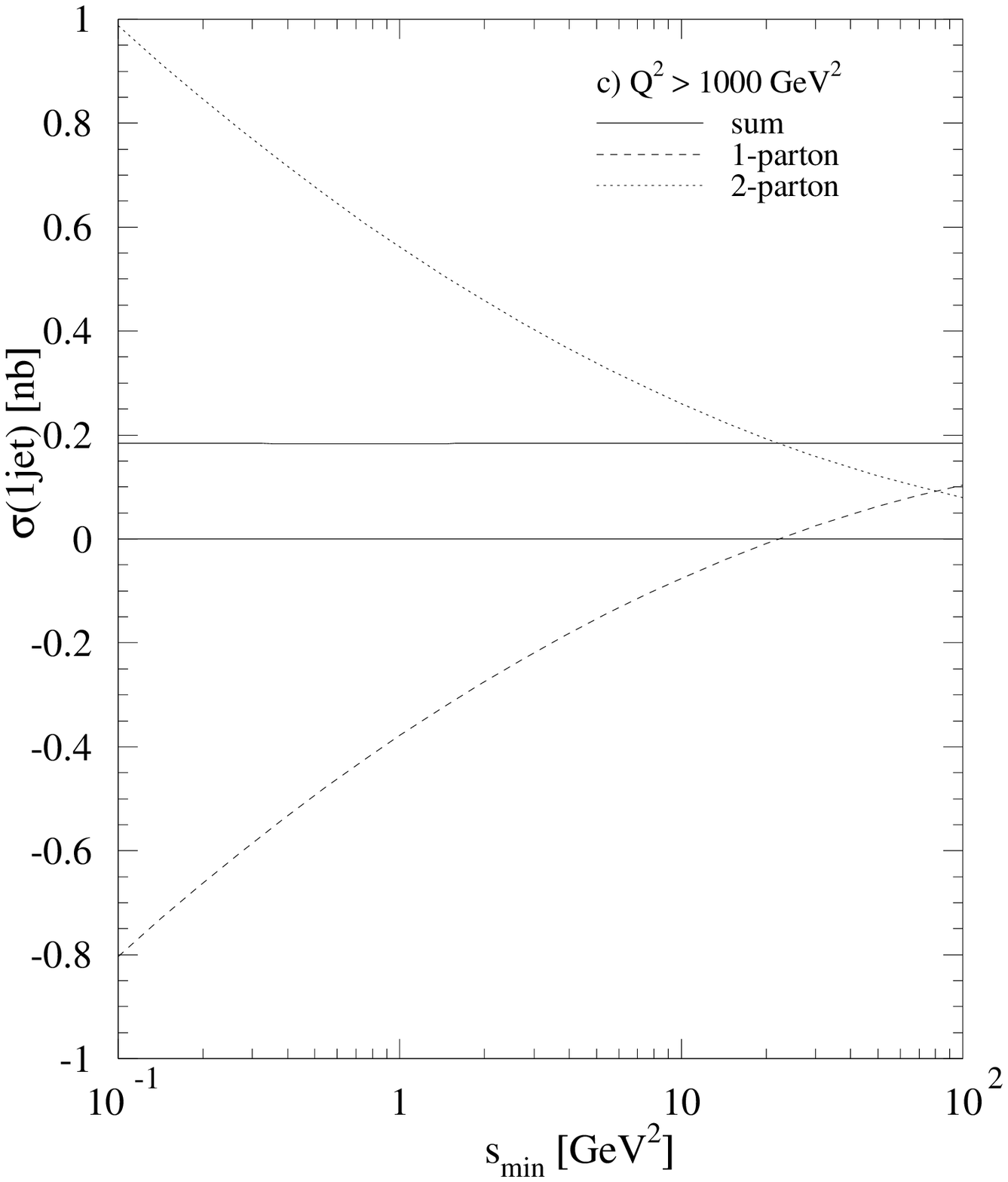,width=8cm}}
    \put(83,0){\epsfig{file=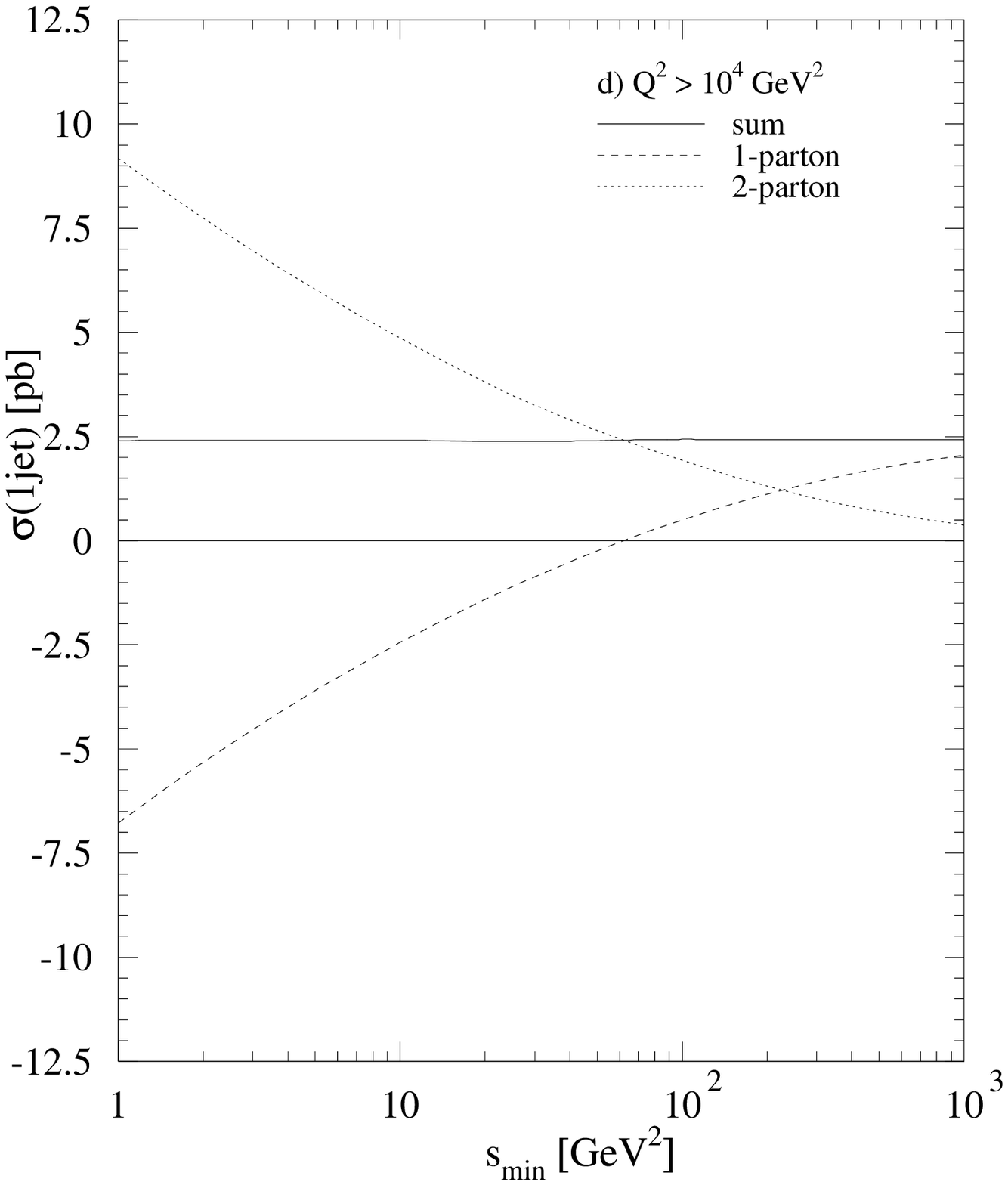,width=8cm}}
  \end{picture}
\caption{Inclusive single-jet cross section for
  $E_T^{\mbox{\scriptsize lab}}>5$~GeV, $|\eta^{\mbox{\scriptsize
  lab}}|<2$ and $Q^2>10$~GeV$^2$ as a function of $\smin$. The $\smin$
  values at which the one-parton contributions vanish lie well within
  the $\smin$ independent region for all $Q^2$.\label{f2}}
\end{figure}

In Fig.~\ref{f2} we plot the NLO cross sections for the one-parton final
states, which include the Born term, the virtual corrections and the
soft and collinear contributions, together with the hard two-parton
final states and their sum as a function of $\smin$ for four different
$Q^2$ regions, integrated over the whole $x$-range. In Fig.~\ref{f2}~a we
see how the logarithmic $\smin$ dependence of the two-parton final
states is compensated by the one-parton final states to give an
$\smin$ independent result of  $\sigma = 11.66\pm 0.02$~nb up to values
of $\smin \simeq 10$~GeV$^2$. Above that value a slight variation of
the sum can be observed and the $\smin$ independence is no longer
ensured. For $\smin >30$~GeV$^2$ the one-parton final state obviously
fails to give a correct $\smin$ dependence and the sum of one- and
two-parton final states strongly decreases. We note that at even
larger $\smin$ values, the one-parton final states will again give
zero, which is the second solution of (\ref{step}) which we rejected
in (\ref{master}). As an important result, one also sees that the
value of $\smin$, for which the one-parton final states vanish and the 
two-parton final states give the full answer is well within the
$\smin$ independent region. Indeed, after we have introduced our
$\sminn$-function into the {\tt MEPJET} program we found that the
one-parton final states did give zero and as a result for the
two-parton final state we found $\sigma = 11.59\pm 0.01$~nb, which agrees
with the answer for small $\smin$ very well. Similar results hold for
the larger $Q^2$ ranges, Fig.~\ref{f2}~b--d. The point at which the NLO
one-parton final state contributions vanish are well within the
$\smin$ independent region. This holds also for the largest $Q^2$
values, where the absolute size of the $\smin$ function is rather
large of the order of $100$~GeV$^2$, as we have seen in Fig.~\ref{f1}. At
the largest $Q^2$ values the results seem to become even more stable with
respect to the $\smin$ dependence.

For a more detailed study we have calculated the single-jet inclusive
cross section for nine different bins in $x$ and $Q^2$, namely $Q^2\in
[10,10^2]$~GeV$^2$, $Q^2\in [10^2,10^3]$~GeV$^2$ and 
$Q^2\in [10^3,10^4]$~GeV$^2$ together with $x\in [10^{-3},10^{-2}]$, 
$x\in [10^{-2},0.1]$ and finally $x\in [0.1,1]$. The actual bins are
summarized in Tab.~\ref{tab1}. In addition, we have
tested the scale dependence by varying the squared renormalization and
factorization scales together by a factor of 4, i.e., $\mu^2 =
\mu_R^2=\mu_F^2 = \xi Q^2$ with $\xi = \frac{1}{4},1,4$. The results
are shown in Tab.~\ref{tab2}--\ref{tab4} in pb, also indicating the
relative difference 
$\Delta=|\sigma_{\mboxsc{std}}-\sigma_{\mboxsc{mod}}|/\sigma_{\mboxsc{std}}$
of the standard PSS method to the modified
PSS. For all $Q^2$ intervals we find agreement of our modified
approach compared to the $\smin$ independent standard approach to
around one percent or better. The overall scale dependence is small,
indicating a very good perturbative stability, as to be
expected. However, it was not our intention to test the scale
dependence, but to test whether our $\sminn$ function would reproduce
the scale behaviour correctly.

This concludes our numerical studies, showing the equivalence of the
standard PSS method with our modified approach by
integrating out only the two-parton final states down to a dynamical
$\sminn$ function given by eqn~(\ref{master}).

\section{Summary and outlook}

We have given a prescription for combining fixed NLO matrix elements
with PS's within the PSS method. It consists in removing the Born,
virtual and soft/collinear contributions from the NLO cross section
for the $n$-jet region by adjusting the PSS parameter $\smin$ for 
each phase space point and each scale $\mu_R,\mu_F$. These
contributions are then included in the hard part of the NLO matrix
elements, which are positive definite. This allows to directly
redistribute the weight provided from these matrix elements with a PS
algorithm.

For the case of inclusive single-jet production in $eP$-scattering 
at ${\cal O}(\alpha_s)$ we have calculated the dynamical $\smin$
parameter for each phase space point $x$ and $Q^2$ and each scale
$\mu_R,\mu_F$. We have numerically compared the standard calculation
with our new approach of evaluating fixed NLO contributions and found
the new approach to give reliable results. We especially found that
the values of $\smin$, for which the virtual plus soft/collinear
contributions vanish, are small enough for the soft and collinear
approximations, used in in the PSS method, to be valid. We note that
the cut-off function has been successfully implemented in the 
{\tt RAPGAP} event generator \cite{rapgap}, which includes the 
${\cal O}(\alpha_s)$ tree level matrix elements. Numerical results and
comparison to data will be discussed in a forthcoming paper. 

The next, more complicated step is the case of dijet production 
in $eP$-scattering, which is especially interesting 
because it allows a precise determination of $\alpha_s$ or
the gluon density in the proton. NLO calculations in the PSS method
from which the cut-off function can be determined are available
\cite{Gr,mepjet,mirkes,fxnlo1}. It might turn out that the PSS method
has to be suplemented with the hybrid method to numerically evaluate
terms of order $\smin\ln(\smin)$ as outlined in section 2. NLO
calculations within the subtraction method are available for dijet
production in $eP$-scattering \cite{CS}, so that the expressions
needed for the hybrid of PSS and subtraction method can readily be
evaluated. We finally note that the ${\cal O}(\alpha_s^2)$ tree level
matrix elements for $eP$-scattering interfaced with the PS are not yet
available in a working Monte Carlo event generator.

\acknowledgments
We have benefited from discussions with S.~Catani, H.~Jung, G.~Kramer,
J.~Rathsman, T.~Sch\"orner, T.~Sj\"ostrand and M.~Sutton. I am grateful to
G.~Kramer for comments on the manuscript. D.~Chapin and N.~Kauer have
been helpful concerning details of the {\tt MEPJET} program.

\begin{appendix}
\section{Crossing functions}

For reasons of completeness, in this appendix we collect from
\cite{mirkes} the definitions of the functions $A_{p\to a}$ and
$B_{p\to a}^{{\overline{\mboxsc{MS}}}}$ which are needed to compute
the crossing functions $C_i^{\overline{\mboxsc{MS}}}$.
The functions are defined via  a one dimensional integration over the
parton densities $f_p$, which also involves the integration over
$()_+$ prescriptions. The finite, scheme independent 
functions $A_{p\to a}(x,\mu_F)$ are given by:
\begin{eqnarray}
A_{g\rightarrow g} &=& \int_x^1\frac{dz}{z}\,\, f_g(x/z,\mu_F)
\left\{\frac{(11N_C-2n_f)}{6N_C}\delta(1-z)\right. \nonumber\\
&&\left.
\hspace{2cm}\,+ \, 
2\left( \frac{z}{(1-z)_+}+\frac{(1-z)}{z}+z(1-z)\right)\right\}
\label{aggdef}
\\
A_{q\rightarrow q} &=& \int_x^1 \frac{dz}{z}\,\,f_q(x/z,\mu_F)
\frac{2C_F}{3}\,\left\{\frac{3}{4}\delta(1-z)\,+\,
\frac{1}{2}\left(\frac{1+z^2}{(1-z)_+}\right)\right\}
\label{aqq}
\\
A_{g\rightarrow q} &=& \int_x^1 \frac{dz}{z}\,\, f_g(x/z,\mu_F)
\,\,\,\frac{1}{4}\,\,\, 
\hat{P}^{(4)}_{g\rightarrow q}(z)
\label{agq}
\\
A_{q\rightarrow g} &=& \int_x^1 \frac{dz}{z}\,\, f_q(x/z,\mu_F)
\,\,\,\frac{1}{4}\,\,\,
\hat{P}^{(4)}_{q\rightarrow g}(z)
\end{eqnarray}
The scheme dependent functions 
$B_{p\rightarrow h}^{{\overline{\mboxsc{MS}}}}(x,\mu_F)$ are given by:
\begin{eqnarray}
B_{g\rightarrow g}^{{\overline{\mboxsc{MS}}}} &=& \int_x^1 \frac{dz}{z}\,\,
f_g(x/z,\mu_F)
\left\{
\left(\frac{\pi^2}{3} -\frac{67}{18}+\frac{5n_f}{9N_C}\right)\delta(1-z)
+2z\left(\frac{\ln(1-z)}{(1-z)}\right)_+ \right.\nonumber\\
& &\hspace{3.5cm}\left.
+2\left( \frac{(1-z)}{z}+z(1-z)\right)\ln(1-z)\right\}
\label{bggdef}
\\
B_{q\rightarrow q}^{{\overline{\mboxsc{MS}}}} &=& \int_x^1 \frac{dz}{z}\,\,
f_q(x/z,\mu_F)
\frac{2C_F}{3}\,\left\{\left(\frac{\pi^2}{6}-\frac{7}{4}\right)\delta(1-z)
+\frac{1}{2}(1-z)\right.
\nonumber\\
& &\hspace{3.5cm}\left.
+\frac{1}{2}(1+z^2)
\left(\frac{\ln(1-z)}{(1-z)}\right)_+\right\}
\label{bqq}
\\
B_{g\rightarrow q}^{{\overline{\mboxsc{MS}}}} &=& \int_x^1 \frac{dz}{z}\,\,
f_g(x/z,\mu_F)
\,\,\,\frac{1}{4}\,\,\,\left\{ 
\hat{P}^{(4)}_{g\rightarrow q}(z)
\ln(1-z)-
\hat{P}^{(\epsilon)}_{g\rightarrow q}(z)\right\}
\label{bgq}
\\
B_{q\rightarrow g}^{{\overline{\mboxsc{MS}}}}
 &=& \int_x^1 \frac{dz}{z}\,\,f_q(x/z,\mu_F)
\,\,\,\frac{1}{4}\,\,\,\left\{ 
\hat{P}^{(4)}_{q\rightarrow g}(z)
\ln(1-z)-
\hat{P}^{(\epsilon)}_{q\rightarrow g}(z)\right\}
\end{eqnarray}
Here, $n_f$ denotes the number of flavors and $N_C=3$ is the number of
colors. The Altarelli-Parisi kernels in the previous equations are
defined by:
\begin{eqnarray}
\hat{P}^{(n\neq 4)}_{g\rightarrow g}(z) &=& 
{P}^{(n\neq 4)}_{g\rightarrow g}(z)\,\,\,\, =
\,4\,
\left( \frac{z}{1-z}+\frac{1-z}{z}+z(1-z)  \right) \\
\hat{P}^{(n\neq 4)}_{q\rightarrow g}(z) &=& 
\frac{8}{9}
{P}^{(n\neq 4)}_{q\rightarrow g}(z) =
\frac{16}{9}
\left( \frac{1+(1-z)^2}{z}-\epsilon z  \right) \\
\hat{P}^{(n\neq 4)}_{g\rightarrow q}(z) &=& 
\frac{1}{3}
P^{(n\neq 4)}_{g\rightarrow q}(z) = 
\frac{2}{3}\,\left( \frac{z^2+(1-z)^2-\epsilon}{1-\epsilon}
                      \right)  \label{pgqdef}\\
\hat{P}^{(n\neq 4)}_{q\rightarrow q}(z) &=& 
\frac{8}{9}
P^{(n\neq 4)}_{q\rightarrow q}(z) = 
\frac{16}{9}
\left( \frac{1+z^2}{1-z}-\epsilon(1-z)   \right)  \label{pqqdef}
\end{eqnarray}
The $P_{ij}^{(\epsilon)}$ are the $\epsilon$ dimensional part
of these $n-$dimensional splitting functions
\begin{eqnarray}
\hat{P}^{(\epsilon)}_{q\rightarrow g}(z) &=& 
\frac{8}{9}
{P}^{(\epsilon)}_{q\rightarrow g}(z) =
-\frac{8}{9}\,2z \\[2mm]
\hat{P}^{(\epsilon)}_{g\rightarrow q}(z) &=& 
\frac{1}{3}
P^{(\epsilon)}_{g\rightarrow q}(z) = 
-\frac{4}{3}\,z(1-z)      \label{pgqepsi}   
\end{eqnarray}
The $()_+$ prescriptions in these equations are defined for an
arbitrary test function $G(z)$ (which is well behaved at $z=1$) as
\bq
 \int\limits_x^1 dz F_+(z)G(z) = \int\limits_x^1 dz
  F(z)[G(z)-G(1)] + G(1)\int\limits_0^x dz F(z) 
  \label{plusdef} 
\eq
The structure and use of the crossing functions are completely analog
to the usual parton distribution function. 

The numerical integrations have been performed in a computer program,
which is provided together with the fixed order Monte Carlo program 
{\tt MEPJET} \cite{mepjet}. The results for $A_{p\to a}$ and 
$B_{p\to a}^{{\overline{\mboxsc{MS}}}}$  for different values of $x$ and
$\mu_F$ are stored in an array in complete analogy to the usual parton
densities, which allows a convenient and numerically quick evaluation
of the crossing functions. 

\end{appendix}


\begin{footnotesize}
\begin{table}
\begin{center}
\begin{tabular}[h]{lccc}
  \multicolumn{1}{c}{\rule[-2.5mm]{0mm}{8mm}}
  & \makebox[3.1cm]{$Q^2 \in [10,10^2]$}
  & \makebox[3.1cm]{$Q^2 \in [10^2,10^3]$}
  & \makebox[3.1cm]{$Q^2 \in [10^3,10^4]$}
\\ \hline
$x \in [10^{-1},1]$ \rule[-2.5mm]{0mm}{8mm}
  & \makebox[1.2cm]{---}
  & \makebox[1.2cm]{bin 3}
  & \makebox[1.2cm]{bin 6}
\\ 
$x \in [10^{-1},10^{-2}]$ \rule[-2.5mm]{0mm}{8mm}
  & \makebox[1.2cm]{bin 1}
  & \makebox[1.2cm]{bin 4}
  & \makebox[1.2cm]{bin 7}
\\ 
$x \in [10^{-2},10^{-3}]$ \rule[-2.5mm]{0mm}{8mm}
  & \makebox[1.2cm]{bin 2}
  & \makebox[1.2cm]{bin 5}
  & \makebox[1.2cm]{---} \\
\end{tabular}
\end{center}
\caption{\label{tab1} Bins in $x$ and $Q^2$ (given in
  GeV$^2$) for $y>0.04$. Two bins are kinematically excluded.}
\end{table}
\end{footnotesize}

\begin{table}
\begin{center}
\begin{tabular}[h]{cccc}
  \multicolumn{4}{c}{Scale $\mu^2=Q^2$} \\ \hline
  bin
  & \makebox[2.2cm]{standard PSS}
  & \makebox[2.2cm]{modified PSS}  
  & $\Delta$ \\ \hline
1\rule[-2.5mm]{0mm}{8mm}
  & \pmdg{1044}{3}
  & \pmdg{1037}{2}
  & 0.7 \\ 
2\rule[-2.5mm]{0mm}{8mm}
  & \pmdg{5734}{15}
  & \pmdg{5700}{9}
  & 0.6 \\ 
3\rule[-2.5mm]{0mm}{8mm}
  & \pmdg{19.99}{0.18}
  & \pmdg{20.22}{0.04}
  & 1.2 \\ 
4\rule[-2.5mm]{0mm}{8mm}
  & \pmdg{2193}{3}
  & \pmdg{2214}{4}
  & 1.0 \\ 
5\rule[-2.5mm]{0mm}{8mm}
  & \pmdg{1882}{4}
  & \pmdg{1875}{2}
  & 0.4 \\ 
6\rule[-2.5mm]{0mm}{8mm}
  & \pmdg{54.10}{0.20}
  & \pmdg{55.60}{0.18}
  & 2.8 \\
7\rule[-2.5mm]{0mm}{8mm}
  & \pmdg{129.3}{0.5}
  & \pmdg{131.1}{0.4}
  & 1.4
\end{tabular}
\end{center}
\caption{\label{tab2}Cross sections in pb for the bins 1--7
  defined in Tab.~\ref{tab1}, comparing the standard PSS method with the
  $\sminn$ modified PSS method for three different scales. $\Delta$
  gives the difference of both methods in percent.}
\end{table}

\begin{table}
\begin{center}
\begin{tabular}[h]{cccc}
  \multicolumn{4}{c}{Scale $\mu^2=4Q^2$} \\ \hline
  bin
  & \makebox[2.2cm]{standard PSS}
  & \makebox[2.2cm]{modified PSS}  
  & $\Delta$ \\ \hline
1\rule[-2.5mm]{0mm}{8mm}
  & \pmdg{1124}{4}
  & \pmdg{1120}{2}
  & 0.4 \\ 
2\rule[-2.5mm]{0mm}{8mm}
  & \pmdg{6218}{15}
  & \pmdg{6191}{7}
  & 0.4 \\ 
3\rule[-2.5mm]{0mm}{8mm}
  & \pmdg{19.97}{0.18}
  & \pmdg{20.15}{0.04}
  & 1.0 \\ 
4\rule[-2.5mm]{0mm}{8mm}
  & \pmdg{2260}{4}
  & \pmdg{2288}{5}
  & 1.2 \\ 
5\rule[-2.5mm]{0mm}{8mm}
  & \pmdg{1987}{6}
  & \pmdg{1981}{2}
  & 0.3 \\
6\rule[-2.5mm]{0mm}{8mm}
  & \pmdg{54.21}{0.22}
  & \pmdg{55.69}{0.13}
  & 2.7 \\ 
7\rule[-2.5mm]{0mm}{8mm}
  & \pmdg{130.9}{0.8}
  & \pmdg{132.1}{0.2}
  & 1.0
\end{tabular}
\end{center}
\caption{\label{tab3}Same as Tab.~\ref{tab2} for $\mu^2=4Q^2$.}
\end{table}

\begin{table}
\begin{center}
\begin{tabular}[h]{cccc}
  \multicolumn{4}{c}{Scale $\mu^2=\mbox{$\frac14$}Q^2$} \\ \hline
  bin
  & \makebox[2.2cm]{standard PSS}
  & \makebox[2.2cm]{modified PSS}  
  & $\Delta$ \\ \hline
1\rule[-2.5mm]{0mm}{8mm}
  & \pmdg{984.1}{3}
  & \pmdg{970.7}{2}
  & 1.4 \\ 
2\rule[-2.5mm]{0mm}{8mm}
  & \pmdg{5412}{14}
  & \pmdg{5343}{6}
  & 1.3 \\ 
3\rule[-2.5mm]{0mm}{8mm}
  & \pmdg{20.00}{0.21}
  & \pmdg{20.30}{0.05}
  & 1.5 \\ 
4\rule[-2.5mm]{0mm}{8mm}
  & \pmdg{2134}{5}
  & \pmdg{2158}{6}
  & 1.1 \\ 
5\rule[-2.5mm]{0mm}{8mm}
  & \pmdg{1781}{6}
  & \pmdg{1758}{2}
  & 1.3 \\ 
6\rule[-2.5mm]{0mm}{8mm}
  & \pmdg{54.90}{0.29}
  & \pmdg{55.90}{0.21}
  & 1.8 \\ 
7\rule[-2.5mm]{0mm}{8mm}
  & \pmdg{129.0}{0.9}
  & \pmdg{130.3}{0.3}
  & 1.0 
\end{tabular}
\end{center}
\caption{\label{tab4}Same as Tab.~\ref{tab2} for $\mu^2=\mbox{$\frac14$}Q^2$.}
\end{table}

\end{document}